\documentclass[prb,twocolumn,showkeys,showpacs,amsmath,amsfonts,floatfix,superscriptaddress,nofootinbib]{revtex4-1}

\usepackage{amssymb,amsmath,amstext}                
\usepackage{graphicx}                                               
\usepackage{epstopdf}                                               
\usepackage{color}       
\usepackage{bm}
\usepackage{appendix}                                              
\usepackage[latin2]{inputenc}
\usepackage{ulem}
\usepackage{latexsym}
\usepackage[colorlinks=true,citecolor=blue,linkcolor=magenta]{hyperref}
\def\be{\begin{equation}}
\def\ee{\end{equation}}
\def\bea{\begin{eqnarray}}
\def\eea{\end{eqnarray}}
\def\bi{\begin{itemize}}
\def\ei{\end{itemize}}

\newcommand{\bra}[1]{\mbox{$\langle #1 |$}}
\newcommand{\ket}[1]{\mbox{$| #1 \rangle$}}

\begin{document}

\title{ Biased dynamics of the miscible-immiscible quantum phase transition 
        in a binary Bose-Einstein condensate }

\author{Francis A. Bayocboc, Jr.} 
\affiliation{Jagiellonian University, 
             Faculty of Physics, Astronomy and Applied Computer Science,
             Institute of Theoretical Physics, 
             ul. \L{}ojasiewicza 11, 30-348 Krak\'ow, Poland }  
\affiliation{Jagiellonian University, 
             Mark Kac Center for Complex Systems Research,
             ul. \L{}ojasiewicza 11, 30-348 Krak\'ow, Poland }  
 
\author{Jacek Dziarmaga} 
\affiliation{Jagiellonian University, 
             Faculty of Physics, Astronomy and Applied Computer Science,
             Institute of Theoretical Physics, 
             ul. \L{}ojasiewicza 11, 30-348 Krak\'ow, Poland }  
\affiliation{Jagiellonian University, 
             Mark Kac Center for Complex Systems Research,
             ul. \L{}ojasiewicza 11, 30-348 Krak\'ow, Poland }  

\author{Wojciech H. Zurek}
\affiliation{Theory Division, Los Alamos National Laboratory, Los Alamos, New Mexico 87545, USA}       

\date{\today}

\begin{abstract}
A quantum phase transition from the miscible to the immiscible phase of a quasi-one-dimensional binary Bose-Einstein condensate is driven by ramping down the coupling amplitude of its two hyperfine states. It results in a random pattern of spatial domains where the symmetry is broken separated by defects. In distinction to previous studies [J. Sabbatini et al., Phys. Rev. Lett. 107, 230402 (2011), New J. Phys. 14 095030 (2012)], we include nonzero detuning between the light field and the energy difference of the states, which provides a bias towards one of the states.
Using the truncated Wigner method, we test the biased version of the quantum Kibble-Zurek mechanism [M. Rams et al., Phys. Rev. Lett. 123, 130603 (2019)] and observe a crossover to the adiabatic regime when the quench is sufficiently fast to dominate the effect of the bias. 
We verify a universal power law for the population imbalance in the nonadiabatic regime both at the critical point and by the end of the ramp.
Shrinking and annihilation of domains of the unfavourable phase after the ramp, that is, already in the broken symmetry phase, enlarges the defect-free sections by the end of the ramp.
The consequences of this phase-ordering effect can be captured by a phenomenological power law. 
\end{abstract}

\maketitle

\section{Introduction}
\label{sec:intro}

Quantum phase transitions involve a dramatic change in the ground state of the system as a consequence of small changes to its Hamiltonian. They can be induced by adjusting an external parameter such as magnetic field. They need not happen at absolute zero temperature: it is sufficient that the temperature is sufficiently low for the measurable equilibrium properties of the system (e.g., correlations) to be dominated by the properties of the ground state. The miscibility-immiscibility transition in the Bose-Einstein condensate (BEC) is a good illustration of a quantum phase transition. 

\begin{figure}[t!]
\vspace{-0cm}
\includegraphics[width=0.999\columnwidth,clip=true]{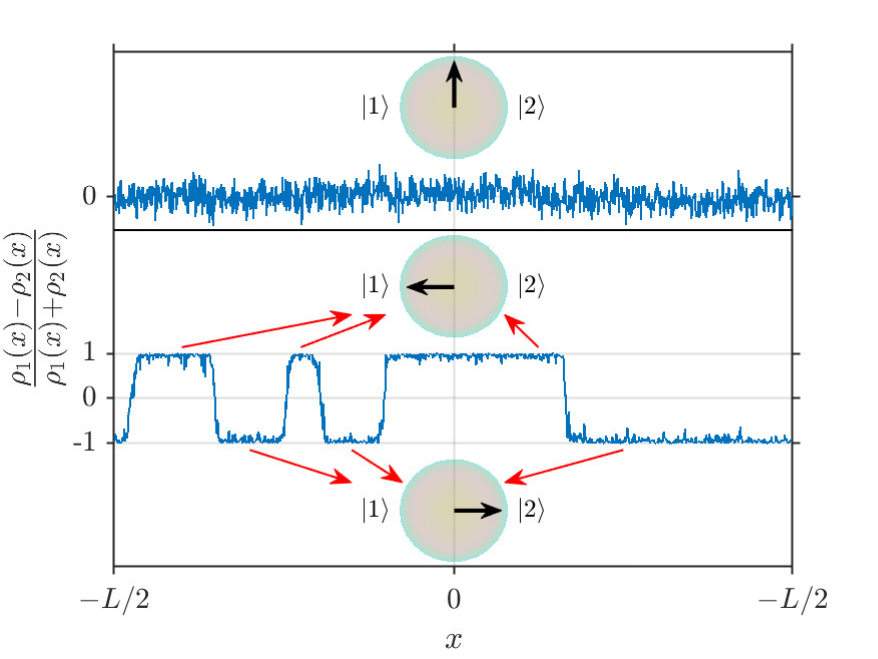}
\vspace{-0cm}
\caption{
{Miscible-immiscible transition. }
A condensate of atoms in an equal superposition of two hyperfine states is driven across a miscible-immiscible transition and separates into domains with different states. A typical size of the domains is proportional to the third root of the transition time \cite{sabbatiniPRL,sabbatiniNJP}.
In this paper, we apply a bias favoring one of the states and study how it affects the outcome of the transition.
}
\label{fig:bloch}
\end{figure}

For illustration, atoms of the condensate (such as ${}^{87}$Rb) may start in a superposition of two hyperfine states. In the presence of the magnetic field, these states are miscible, so these atoms persist in superposition. However, as the field is lowered, hyperfine states of ${}^{87}$Rb become immiscible, inducing symmetry breaking: different BEC fragments attempt to choose one or the other of these two hyperfine states  (see
Fig. \ref{fig:bloch} for an example of such a transition). By controlling an external parameter, one can drive BEC atoms through such a miscibility-immiscibility transition at various rates.

The miscibility-immiscibility transition is in some ways reminiscent of the paramagnetic-ferromagnetic transition in the quantum Ising chains in a transverse field in that the system is forced to choose between the two possible alternatives---spins up or down in the ferromagnetic phase of the Ising model and one or the other of the two hyperfine states in the immiscible phase of the BEC. We, therefore, expect that the Kibble-Zurek mechanism (KZM) that has been by now well established in the other phase transitions can also be studied in the miscibility-immiscibility transitions in the Bose-Einstein condensates \cite{sabbatiniPRL,sabbatiniNJP}.

KZM originated from a scenario for topological defect formation in cosmological phase transitions driven by expanding universe~\cite{K-a, *K-b, *K-c} where independent selection of broken symmetry vacua in causally disconnected regions can be expected to result in a mosaic of broken symmetry domains leading to topologically nontrivial configurations. 
However, for phase transitions in condensed matter systems, relativistic causality is not relevant. Thus, to relate the density of defects to the quench rate and the nature of the transition, a dynamical theory for the continuous phase transitions was proposed.~\cite{Z-a,*Z-b,*Z-c,Z-d} It predicts the scaling of the defect density as a function of the quench rate by employing the universality class of the transitions---its equilibrium critical exponents. It has been verified by numerous simulations~\cite{KZnum-a,KZnum-b,KZnum-c,KZnum-d,KZnum-e,KZnum-f,KZnum-g,*KZnum-h,*KZnum-i,KZnum-j,KZnum-k,KZnum-l,KZnum-m,that} and experiments~\cite{KZexp-a,KZexp-b,KZexp-c,KZexp-d,KZexp-e,KZexp-f,KZexp-g,QKZexp-a,KZexp-h,KZexp-i,KZexp-j,KZexp-k,KZexp-l,KZexp-m,KZexp-n,KZexp-o,KZexp-p,KZexp-q,lamporesi2013,donadello2016,KZexp-s,KZexp-t,KZexp-u,KZexp-v,KZexp-w,KZexp-x}. Topological defects play central role in these studies as they can survive inevitable dissipation and can be counted afterwards.

The quantum version of KZM (QKZM) was developed for quenches across critical points in isolated quantum systems ~\cite{QKZ1,QKZ2,QKZ3,d2005,d2010-a, d2010-b, QKZteor-a,QKZteor-b,QKZteor-c,QKZteor-d,QKZteor-e,QKZteor-f,QKZteor-g,QKZteor-h,QKZteor-i,QKZteor-j,QKZteor-k,QKZteor-l,QKZteor-m,QKZteor-n,QKZteor-o,KZLR1,KZLR2,QKZteor-oo,delcampostatistics,KZLR3,QKZteor-q,QKZteor-r,QKZteor-s,QKZteor-t,sonic,QKZteor-u,QKZteor-v,QKZteor-w,QKZteor-x,roychowdhury2020dynamics,sonic, schmitt2021quantum,RadekNowak,dziarmaga_kinks_2022,oscillations}. It was already tested by experiments~\cite{QKZexp-a, QKZexp-b, QKZexp-c, QKZexp-d, QKZexp-e, QKZexp-f, QKZexp-g,deMarco2,Lukin18,adolfodwave,2dkzdwave,King_Dwave1d_2022,Semeghini2021,Satzinger2021etal}. Recent progress in Rydberg atoms' versatile emulation of quantum many-body systems  ~\cite{rydberg2d1,rydberg2d2,Semeghini2021,Satzinger2021etal} and coherent D-Wave~\cite{King_Dwave1d_2022,King_Dwave_glass} open the possibility to study the QKZM in a variety of two- and three-dimensional settings and/or to employ it as a test of quantumness of the hardware~\cite{RadekNowak, King_Dwave1d_2022, dziarmaga_kinks_2022, schmitt2021quantum, oscillations}.

The QKZM can be briefly outlined as follows. 
A smooth ramp crossing the critical point at time $t=0$ can be linearized in its vicinity as 
\begin{equation}
\epsilon(t)=\frac{t}{\tau_Q}.
\label{epsilont}
\end{equation}
Here, $\epsilon$ is a dimensionless parameter in a Hamiltonian that measures distance from the quantum critical point, and $\tau_Q$ is called a quench time. Initially, the system is prepared in its ground state far from the critical point. At first, far from the critical point, the evolution adiabatically follows the ground state of the changing Hamiltonian. However, adiabaticity fails near the time $-\hat t$ when the energy gap becomes comparable to the quench ramp rate: \begin{equation}
\Delta\propto|\epsilon|^{z\nu} \propto |\dot \epsilon/\epsilon| = 1/|t|  
\end{equation}
and the critical slowing down precludes such adiabatic following.

This timescale is 
\be 
\hat t\propto \tau_Q^{z\nu/(1+z\nu)},
\label{hatt}
\ee
where $z$ and $\nu$ are the dynamical and the correlation length critical exponents, respectively. 
The correlation length at $-\hat t$, 
\begin{equation}
\hat\xi \propto \tau_Q^{\nu/(1+z\nu)}, 
\label{hatxi}
\end{equation}
defines the size of the domains where fluctuations select the same broken symmetry ground state. Its inverse determines the resulting density of defects left after crossing the critical point;
\be 
N_d \propto \hat\xi^{-1}.
\ee
The two KZ scales are related by
\be   
\hat t \propto \hat\xi^z.
\ee 
Accordingly, in the KZM regime after $-\hat t$, observables are expected to satisfy the KZM dynamical scaling hypothesis~\cite{KZscaling1,KZscaling2,Francuzetal} with $\hat\xi$ being the unique scale. For, say, a two-point observable ${\cal O}_r$, where $r$ is a distance between the two points, it reads
\be 
\hat\xi^{\Delta_{\cal O}} \bra{\psi(t)} {\cal O}_r \ket{\psi(t)} = 
F_{\cal O}\left( t/\hat\xi^z , r/\hat\xi \right),
\label{eq:KZscalingO}
\ee
where $\ket{\psi(t)}$ is the state during the quench, $\Delta_{\cal O}$ is the scaling dimension, and $F_{\cal O}$ is a non-universal scaling function. 

\section{Quench with a bias}
\label{sec:bias}

The selection of the broken symmetry can be biased and, simultaneously, the quantum transition can be made more adiabatic, by adding a bias term to the Hamiltonian that is linear in the order parameter with a bias strength $b$ \cite{QKZteor-r}. A similar mechanism was demonstrated experimentally for a classical thermodynamic transition in helium-3 \cite{KZexp-x}. In a quantum transition, the bias opens a finite energy gap at the critical point,
\be
\Delta_b \propto b^{z\nu/(\beta\delta)},
\label{eq:Deltab}
\ee
and makes the correlation length finite:
\be 
\xi_b \propto \Delta_b^{-1/z} \propto b^{-\nu/(\beta\delta)}.
\label{eq:xib}
\ee
Here $\beta$ is the order parameter exponent in the ordered phase ($M\propto \epsilon^\beta$, where $M$ is the order parameter) and $\delta$ is its exponent at the critical point ($M\propto b^{1/\delta}$).  
With $\xi_b$ providing an additional length scale, the scaling hypothesis \eqref{eq:KZscalingO} generalizes to
\be 
\hat\xi^{\Delta_{\cal O}} \bra{\psi(t)} {\cal O}_r \ket{\psi(t)} = 
F_{\cal O}\left( t/\hat\xi^z , \hat\xi/\xi_b, r/\hat\xi \right).
\label{eq:KZscalingb}
\ee
The extra argument, $\hat\xi/\xi_b$, discriminates between the non-adiabatic and adiabatic regimes. 
When $\hat\xi\gg\xi_b$, the energy gap \eqref{eq:Deltab} is strong enough to make the quench adiabatic all the way through the critical point. 
When $\hat\xi\ll\xi_b$, then in first approximation, the bias can be ignored and the QKZM proceeds as usual. The freeze-out takes place far enough from the critical point for the weak bias to have a negligible effect. 
Beyond this first approximation, one can expect that, before $-\hat t$, when the evolution is adiabatic, the order parameter in the ground state is proportional to $|\epsilon|^{-\gamma}b$.
Here $|\epsilon|^{-\gamma}$ is proportional to the linear susceptibility and $\gamma$ is the susceptibility exponent. At $-\hat t$, it freezes out with a value proportional to
\be 
\hat M \propto b ~\tau_Q^{\gamma/(1+z\nu)}.
\ee
This is the order parameter when the system is crossing the critical point. It remains a non-universal system-specific question of whether this characteristic power law survives after the quench deep in the symmetry-broken phase. 

\section{System}
\label{sec:system}

In this paper, we consider the effect of the bias on the miscibility-immiscibility transition in the same system as in Refs. \onlinecite{sabbatiniPRL,sabbatiniNJP}. The Hamiltonian for the binary BEC mixture in one dimension reads \cite{lee2009,gligoric2010} 
\begin{equation}
\hat{H} =  \hat{H}_{sp} + \hat{H}_{\rm int} + \hat{H}_{\rm cpl}.
\label{H}
\end{equation}
Here $\hat{H}_{sp}$, $\hat{H}_{\rm int}$, and $\hat{H}_{\rm cpl}$ are the single-particle, interaction, and coupling Hamiltonians, respectively, defined as
\begin{eqnarray}
\hat{H}_{sp} & =  & \int dx \sum_{i=1}^{2}\hat{\psi}_{i}^{\dagger}(x)
\left[ 
-\frac{\hbar^2}{2m}\frac{\partial^2}{\partial x^2}-\mu+V(x)
\right]\hat{\psi}_{i}(x),\\
\hat{H}_{\rm int} & = & \int dx\left\{ \sum_{i=1}^{2}\frac{g_{ii}}{2}\hat{\psi}_{i}^{\dagger}(x)\hat{\psi}_{i}^{\dagger}(x)\hat{\psi}_{i}(x)\hat{\psi}_{i}(x)\right.\nonumber\\
 & & \left. + g_{12}\hat{\psi}_{1}^{\dagger}(x)\hat{\psi}_{2}^{\dagger}(x)\hat{\psi}_{2}(x)\hat{\psi}_{1}(x)\right\},
\label{eq:interactionHam}\\
\hat{H}_{\rm cpl} & = & 
\int dx
\left\{
\frac{\hbar b}{2} 
\left[
\hat{\psi}_{2}^{\dagger}(x)\hat{\psi}_{2}(x)-
\hat{\psi}_{1}^{\dagger}(x)\hat{\psi}_{1}(x)
\right]
\right.
\nonumber\\
 & & 
 \left. 
 -\hbar \Omega(t) 
 \left[
 \hat{\psi}_{1}^{\dagger}(x)\hat{\psi}_{2}(x) + \hat{\psi}_{2}^{\dagger}(x)\hat{\psi}_{1}(x)
 \right]
 \right\}.
\label{eq:couplingHam}
\end{eqnarray}
Here $\hat{\psi}_{i}(x)$ is the Bose field operator that annihilates a particle in hyperfine state $i$ at position $x$. It obeys $[\hat{\psi}_i(x),\psi^\dag_j(x')] = \delta_{ij}\delta(x-x')$.
$g_{ij}$ are one-dimensional (1D) interaction constants obtained by integration from a 3D Hamiltonian where the transverse state is tightly confined in the transverse ground state by a transverse harmonic potential with frequency $\omega_{\perp}$: $g_{ij} = 2\hbar^2 a_{ij}/(m a_{\perp}^{2})$, where $a_{ij}$ is the 3D \textit{s}-wave scattering length and $a_{\perp} = \sqrt{\hbar/m\omega_{\perp}}$ is the transverse harmonic oscillator length.
In the coupling Hamiltonian, $\Omega(t)$ is the coupling strength and $b$ is the detuning of the light field from the energy difference of the states. The detuning is the bias that favours one of the two components over the other.

In the absence of the bias, $b=0$, the ground state of the model undergoes a continuous phase transition between the miscible phase, when $\Omega>\Omega_c$, and the immiscible one, when $\Omega<\Omega_c$. At the mean-field level, in the former phase, each particle is in a symmetric superposition of the two hyperfine states and in the latter, there are two symmetry-broken ground states where the superposition is tilted in favour of one of the two hyperfine states.
In the following we assume $g_{11}\approx g_{12}\equiv g$ when the critical 
\begin{equation}
    \hbar\Omega_c=\frac12(g_{12}-g)\rho
\label{eq:Omegac}
\end{equation} 
with $\rho$ being total particle density \cite{sabbatiniPRL,sabbatiniNJP}. The linear ramp \eqref{epsilont} is implemented as
\begin{equation}
    \Omega(t)=\Omega_c\left[ 1 - \epsilon(t) \right]
    \label{eq:Omegat}
\end{equation}
starting in the ground state at $2\Omega_c$ and stopping after $\Omega$ is brought down to zero. 


The rest of the paper is organized as follows. In Sec. \ref{sec:model}, we study model \eqref{H} in order to extract all relevant mean-field critical exponents for the miscible-immiscible quantum phase transition. In Sec. \ref{sec:TWA}, we briefly outline the truncated Wigner approximation \cite{blakie2008,steel1998,martin2010,ruostekoski2013BookCh} and anticipate potential problems with the ultraviolet divergence of quantum fluctuations represented by classical ones. The biased QKZM is considered in Secs. \ref{sec:M} and \ref{sec:K}. In Sec. \ref{sec:M}, we focus on the order parameter scaling both when the ramp is crossing the critical point and when it is terminated deep in the immiscible phase. In Sec. \ref{sec:K}, the kinks/defects are counted as a function of the bias driving the QKZM towards a defect-free regime. 
In Sec. \ref{sec:exp}, possible experimental realizations of the model are discussed. 
Finally, we conclude in Sec. \ref{sec:conclusion}.

\section{Model properties}
\label{sec:model}

In the framework of the truncated Wigner approximation (TWA) \cite{blakie2008,steel1998,martin2010,ruostekoski2013BookCh}, the operators in the Hamiltonian \eqref{H} are replaced by classical fields $\psi_i$. In a homogeneous system, $V(x)=0$, the uniform ground state can be parameterized as 
\bea
\psi_1^{(0)} &=& \sqrt{\rho} \cos\left(\frac14\pi-\alpha\right),~~ \nonumber\\
\psi_2^{(0)} &=& \sqrt{\rho} \sin\left(\frac14\pi-\alpha\right).
\label{eq:rho_alpha}
\eea
Here $\rho$ is the total density of particles and $\alpha$ plays a similar role as the order parameter for the miscible-immiscible transition that can be defined as a population imbalance:
\be 
M = \frac{\rho_1-\rho_2}{\rho_1+\rho_2}.
\label{eq:M}
\ee 
Here $\rho_i=|\psi_i|^2$. In the ground state \eqref{eq:rho_alpha} we have $M=\sin 2\alpha$.
The ground state minimizes the energy density
\bea 
\varepsilon(\rho,\alpha) 
&=&
-\mu\rho - 
\frac12\hbar b \rho \sin 2\alpha - 
\hbar\Omega \rho \cos 2\alpha + \nonumber\\
& &
\frac12g\rho^2+\frac14(g_{12}-g)\rho^2 \cos^2 2\alpha.
\label{eq:epsilon}
\eea
Here we assumed $g_{11}=g_{22}\equiv g$ which is a good approximation \cite{sabbatiniPRL,sabbatiniNJP}. 
A minimization with respect to $\rho$ yields a compact formula for the chemical potential,
\be 
\mu=\frac12\rho\left(g_{12}+g\right)-\frac{\hbar\Omega}{\cos2\alpha},
\label{eq:mu}
\ee 
and with respect to $\alpha$ an equation for $b$:
\be 
b = 
2
\left[ 
\frac{\Omega}{\cos 2\alpha} - \Omega_c
\right] 
\sin 2\alpha.
\ee 
Here $\Omega_c$ is the critical value of $\Omega$ in \eqref{eq:Omegac}.

The Ginzburg expansion of the energy \eqref{eq:epsilon} near $\Omega_c$ in powers of $\alpha$ yields
\be 
\varepsilon=
\varepsilon_0+
\hbar\rho
\left[
b\cdot \alpha + 
2 \left( \Omega-\Omega_c \right) \cdot \alpha^2 + 
\Omega_c \cdot \alpha^4
\right].
\label{eq:GL}
\ee
For zero bias, $b=0$, the symmetric $\alpha=0$ is a solution for any $\Omega$, but it is unstable in the immiscible phase below $\Omega_c$. 
Above $\Omega_c$, when the quartic term is neglected for small enough $b$, there is an approximate solution
\be 
\alpha \approx \frac{b}{4\left( \Omega - \Omega_c \right)},
\ee 
that diverges at the transition with the susceptibility exponent $\gamma=1$.
The quartic term prevents this divergence and allows the order parameter at $\Omega=\Omega_c$ to remain finite:
\be 
\alpha_c=\left(\frac{b}{4\Omega_c}\right)^{1/3}
\label{eq:alphab}
\ee 
with the critical exponent $\delta=3$. 

The expansion \eqref{eq:GL} also provides an insight into small Bogoliubov fluctuations around the uniform ground state solution.
For $b=0$ and when the critical point is approached from above, the quadratic term in \eqref{eq:GL} makes the frequency of small oscillations with wave vector $k=0$ around the ground state, $\alpha=0$, decrease as $(\Omega-\Omega_c)^{1/2} $. This power law implies that the critical exponents satisfy $z\nu=1/2$.
For a nonzero bias and at $\Omega=\Omega_c$, small harmonic oscillations around \eqref{eq:alphab} have a frequency $\propto (b/\Omega_c)^{1/3}$. The exponent $1/3$, that stands for $z\nu/\beta/\delta$, implies $\beta=1/2$.
Finally, a linear dispersion, $\omega\propto k$, at the critical point implies $z=1$ and, consequently, $\nu=1/2$.  
This way, we obtained all critical exponents that are relevant for the biased KZM. They are the mean-field exponents for the Ising universality class.
For a quick reference, we also list here the exact exponents that should be valid asymptotically very close to the critical point: $z=1$, $\nu=1$, $\gamma=7/4$, $\delta=15$, and $\beta=1/8$. In principle, they could be probed by QKZM in the limit of very slow quenches.

\section{Truncated Wigner Approximation}
\label{sec:TWA}

In the truncated Wigner approximation \cite{blakie2008,steel1998,martin2010,ruostekoski2013BookCh} the two fields, $\psi_i(t,x)$, evolve according to the classical coupled Gross-Pitaevski equations (GPE)
\begin{eqnarray}\label{eq:GPE}
i\hbar\frac{\partial {\psi_{i}}}{\partial t} &=& 
\left[ -\frac{\hbar^2}{2m}\frac{\partial^2}{\partial x^2} - \mu + V(x) \right] {\psi}_{i}
\nonumber\\
&& 
+ (-1)^{i} \frac{\hbar b}{2}~\psi_i - \hbar\Omega(t)~ {\psi}_{3-i}
\nonumber \\
&&
+ \left[ g_{ii}|\psi_i|^2 + g_{12}|{\psi}_{3-i}|^2 \right] {\psi}_{i}.
\end{eqnarray}
The simulation starts from the ground state above the critical point at $\Omega=2\Omega_c$ and follows the ramp \eqref{eq:Omegat} down to $\Omega=0$ where the ramp stops. 

The initial ground state is dressed with random fluctuations as
\begin{equation}
\psi_i(x,t_{\rm in}) = 
\psi_i^{(0)} + 
\sum_n [\eta_n u_{i,n}(x) + \eta_n^* v_{i,n}^*(x)].
\label{eq:Bog_expansion}
\end{equation}
Here, index $n$ numbers stationary Bogoliubov modes around the initial state and $\eta_n$ are complex Gaussian noises with correlations $\overline{\eta_{n}^*\eta_m}=\delta_{nm}/2$. In the TWA framework, they represent quantum fluctuations in the initial ground state. Each random initial state is evolved with the GPE \eqref{eq:GPE}. Expectation values of observables are estimated by averaging over the random initial noises. Hereafter, the error bars of the estimates account for the standard error of the mean and indicate a 95\% confidence interval.

The representability of the quantum fluctuations by the classical ones in the TWA has inevitable limitations. For instance, the average density in \eqref{eq:Bog_expansion} is
\be 
\rho_i =
\left| \psi_i^{(0)}  \right|^2 +
\sum_n \frac12 \left( |u_{i,n}(x)|^2 + |v_{i,n}(x)|^2 \right)
\label{eq:rho_TWA}
\ee
while the correct formula for a Bogoliubov vacuum reads
\be 
\rho_i =
\left| \psi_i^{(0)}  \right|^2 + \sum_n |v_{i,n}(x)|^2 .
\label{eq:rho_Bog}
\ee
As in our periodic boundary conditions, the Bogoliubov modes are momentum eigenstates,
\be 
u_{i,n}(x)=U_{i,n}e^{ik_nx},~~
v_{i,n}(x)=V_{i,n}e^{ik_nx},
\ee
for every $n$ we have $|u_{i,n}(x)|^2 \propto |v_{i,n}(x)|^2$. The discrepancy between \eqref{eq:rho_TWA} and \eqref{eq:rho_Bog} is negligible for low-frequency modes, with a wavelength much longer than the healing length, where $|U_{i,n}|\approx |V_{i,n}|$. However, for high-frequency modes, where $|U_{i,n}|\approx 1$ and $|V_{i,n}|\ll1$, there is a dramatic difference. As their coefficients $|V_{i,n}|$ become negligible with increasing frequency, they also have a negligible contribution to the exact formula \eqref{eq:rho_Bog} but at the same time, as their $|U_{i,n}|$ become close to $1$, there is an ultra-violet (UV) divergence in the TWA approximation \eqref{eq:rho_TWA}.

At first sight, the error could be mitigated just by truncating the high-frequency modes from the expansion \eqref{eq:Bog_expansion}.
The question of where exactly to truncate is complicated by the fact that the healing length, and thus the cutoff, depends on $\Omega$. It is small at the initial $2\Omega_c$ and large near the critical point, where it grows up to $\xi_b\propto\left(b/\Omega_c\right)^{-1/3}$. 
In the adiabatic regime, where $\xi_b\ll\hat\xi$, all wavelengths evolve adiabatically and it is $\xi_b$ that sets the cut-off scale at the critical point.
In the complementary nonadiabatic regime, where $\hat\xi\ll\xi_b$, wavelengths shorter than $\hat\xi$ evolve adiabatically and, as they are also much shorter than $\xi_b$, they need to be truncated at the critical point.
Wavelengths much longer than $\hat\xi$ freezeout near $-\hat t$, where $\hat\xi$ is the healing length, and thus they do not require the truncation anywhere between $-\hat t$ and the critical point. Therefore, it is $\hat\xi$ that sets the UV cutoff in the non-adiabatic regime.
In the following, we avoid the truncation while bearing in mind the above discussion. 

For our simulations, we choose to simulate ${}^{87}$Rb atoms in a ring trap of circumference $L = 96\,\mu\mathrm{m}$ with transverse trapping frequency $\omega_{\perp} = 2\pi\times500\,\mathrm{Hz}$ and total number of particles $N_{\mathrm{tot}} = N_{1} + N_{2} = 2\times10^{4}$. We take the 3D \textit{s}-wave scattering lengths to be $a_{11} = a_{22} = a_{12}/2 = 1.325\,\mathrm{nm}$ (from which the interaction strengths $g_{ij}$ can be calculated via $g_{ij} = 2\hbar\omega_{\perp}a_{ij}$, in the absence of confinement induced resonances \cite{olshanii1998}). With these parameters, all energy scales are smaller than the energy of the first excited state of the transverse harmonic trap, \textit{e.g.} $\mu_{0} \approx 9.15\times10^{-32} \mathrm{J} < \hbar\omega_{\perp}$, and our system is well within the one-dimensional regime \cite{moritz2003}. Here, $\mu_{0}$ is the chemical potential of the two components $\mu_{1} = \mu_{2} = \mu_{0}$, when both have the same number of particles and $b = 0$. Once we introduce a nonzero bias $b$, the chemical potentials of the two components are given by $\mu_{1} = \mu_{0} + \hbar b/2$ and $\mu_{2} = \mu_{0} - \hbar b/2$. The average of the chemical potentials $\mu = (\mu_{1} + \mu_{2})/2 = \mu_{0}$ is still a constant. Furthermore, the large ratio between the total number of particles $N$ and the number of simulated Bogoliubov modes $M_{B} = 1024$ ensures the validity of the TWA \cite{blakie2008}. All numerical simulations reported hereafter were performed with the software package XMDS2 \cite{xmds2}.

The parameters presented in the preceding paragraph correspond to the regime where the two components are strongly immiscible with $\Delta \equiv g_{11}g_{22}/g_{12}^{2} = 0.25$. These parameters are chosen such that the system spin healing length $\xi_\mathrm{s} \equiv \hbar/\sqrt{2m\rho g_{s}}$, with $g_\mathrm{s} = (2g_{12} - g_{11} - g_{22})/2$, is relatively short, and leads to both a large number of domains and their straightforward identification.

\begin{figure}[t!]
\vspace{-0cm}
\includegraphics[width=0.999\columnwidth,clip=true]{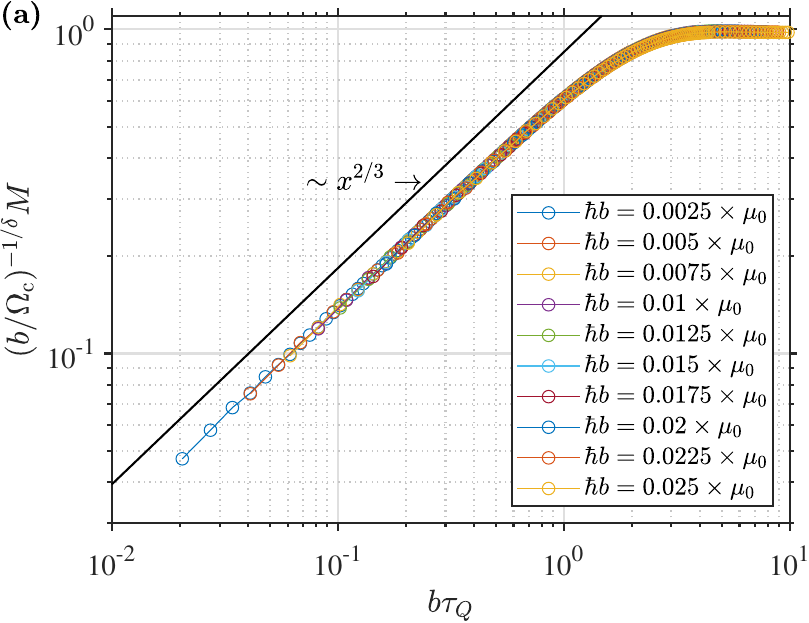}
\includegraphics[width=0.999\columnwidth,clip=true]{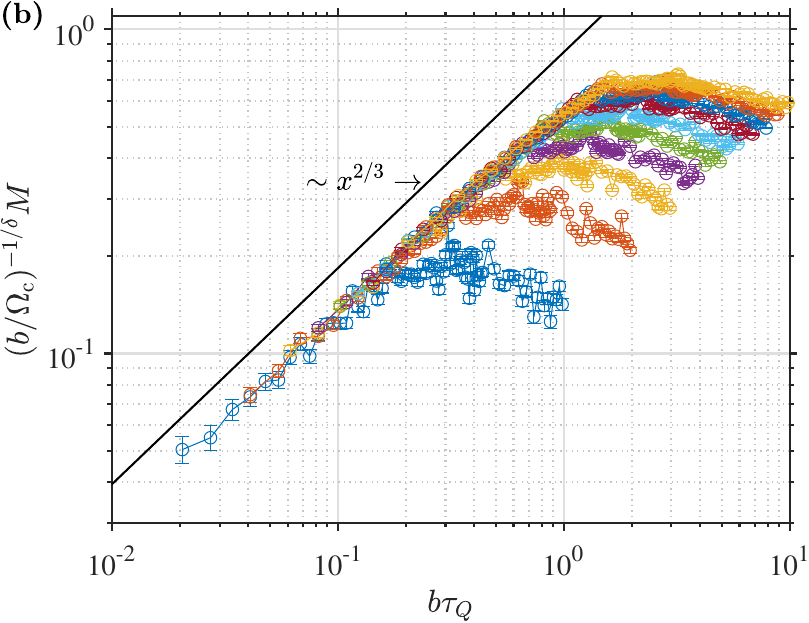}
\vspace{-0cm}
\caption{
{Critical order parameter scaling.}
The order parameter when the ramp is crossing the critical point, at $\Omega=\Omega_c$, as a function of scaled quench time for different biases.
In (a) fluctuations $\eta_n$ in \eqref{eq:Bog_expansion} were set to zero resulting in a perfect collapse in accordance with the scaling hypothesis in \eqref{eq:FMtc} and \eqref{eq:frac23}. We can also see the adiabatic saturation for $b\tau_Q\gg1$. 
In (b) the same but with the classical fluctuations in \eqref{eq:Bog_expansion} and their unphysical UV divergence.
}
\label{fig:collapse_M_c}
\end{figure}

\begin{figure}[t!]
\vspace{-0cm}
\includegraphics[width=0.999\columnwidth,clip=true]{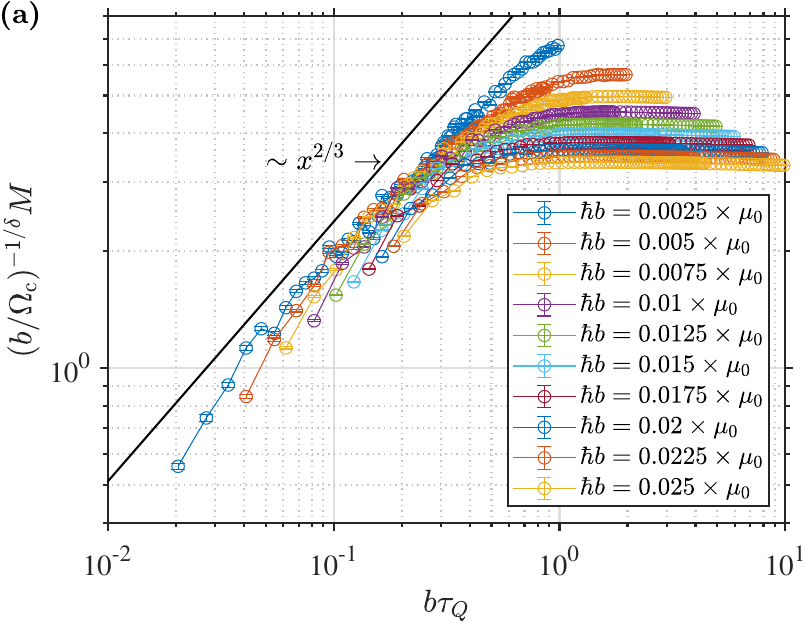}
\includegraphics[width=0.999\columnwidth,clip=true]{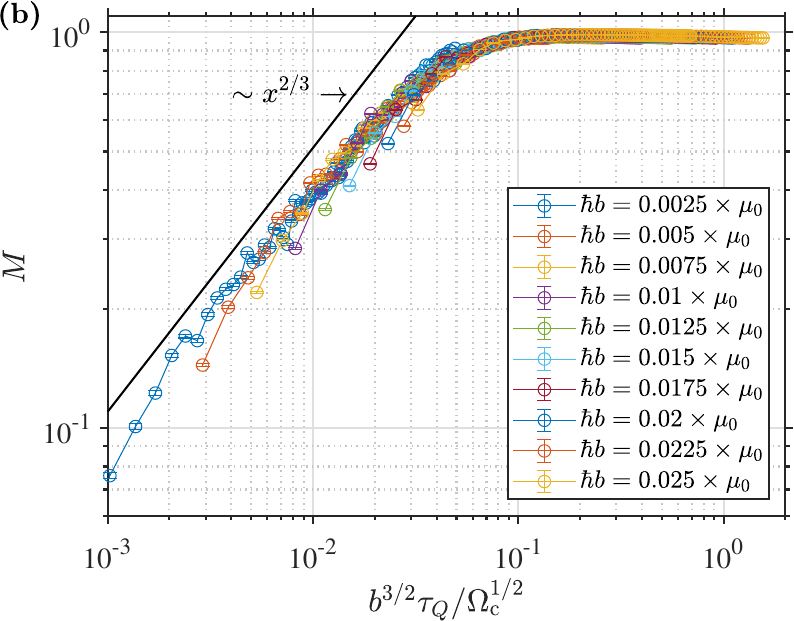}
\vspace{-0cm}
\caption{
{Final order parameter scaling.}
In (a) 
scaled order parameter at the end of the ramp, $\Omega=0$, as a function of scaled quench time $b\tau_Q$ for different biases. For small $b\tau_Q$ the plots collapse in accordance with the scaling hypothesis \eqref{eq:frac23}.
In the adiabatic regime, for large $b\tau_Q$, the order parameter saturates at $1$ for all biases. 
In (b)
the same data as in (a) but presented as the order parameter as a function of $b^{3/2}\tau_Q$.
This scaling makes the plots collapse for both small and large $b^{3/2}\tau_Q$. 
}
\label{fig:collapse_M_f}
\end{figure}

\section{ Order parameter scaling }
\label{sec:M}

According to \eqref{eq:alphab}, in the ground state at the critical point, the order parameter's response to a weak bias is proportional to $\left(b/\Omega_c\right)^{1/\delta}$. Assuming that this parameter sets a scale for magnetization $M$, we can formulate a dynamical scaling hypothesis for the order parameter during the quench between $\pm\hat t$ as \cite{QKZteor-r} 
\be 
\left(b/\Omega_c\right)^{-1/\delta} 
M(t) =  
F_{M}\left[ (t-t_c)/\hat\xi^z , b \tau_Q^{\beta\delta/(1+z\nu)} \right].
\label{eq:FM}
\ee 
Here $F_M$ is a non-universal scaling function. Its first argument is the scaled time measured with respect to time $t_c$ when the critical point is crossed by the ramp. The second one is proportional to $\hat\xi/\xi_b$ in \eqref{eq:KZscalingb}. 
In particular the hypothesis can be probed at the critical point, $t=t_c$, when it predicts that plots for different $b$ of $\left(b/\Omega_c\right)^{-1/\delta}M(t_c)$ as a function of $x=b \tau_Q^{\beta\delta/(1+z\nu)}$ collapse to a common scaling function: 
\be 
\left(b/\Omega_c\right)^{-1/\delta} 
M(t_c) =  
f_{M}\left( b \tau_Q^{\beta\delta/(1+z\nu)} \right).
\label{eq:FMtc}
\ee 
Here $f_M(x)\equiv F_M[0,x]$. 

This function saturates at a constant value in the adiabatic regime, $x\gg1$, where $M(t_c)$ becomes equal to the order parameter in the ground state at the critical point, which is $\propto\left(b/\Omega_c\right)^{1/\delta}$. With $\delta=3$ the mean-field equation \eqref{eq:alphab} implies 
\be 
f_M(x\gg1)\approx2^{1/3}.
\label{eq:fM_large}
\ee 
In the complementary non-adiabatic regime, $x\ll1$, the order parameter is expected to freezeout at $-\hat t$, where it is proportional to $b\hat\epsilon^{-\gamma}\propto b \hat\xi^{\gamma/\nu}\propto b \tau_Q^{\gamma/(1+z\nu)}$, and survive to the critical point as
$
M(t_c) \propto b \tau_Q^{\gamma/(1+z\nu)}.
$ 
Using the scaling relation $\gamma=\beta(\delta-1)$, we can predict 
\cite{QKZteor-r} 
\be 
f_M(x\ll1) \propto 
\left(b\tau_Q\right)^{\gamma\beta^{-1}\delta^{-1}} = 
\left(b\tau_Q\right)^{2/3}
\equiv x^{2/3}.
\label{eq:frac23}
\ee 
In the last equality, we assumed the mean-field exponents. 

The collapse predicted in \eqref{eq:FMtc} and the asymptotes of the scaling function in \eqref{eq:fM_large} and \eqref{eq:frac23} are tested in Fig. \ref{fig:collapse_M_c}.
In its top panel, initial fluctuations $\eta_n$ in \eqref{eq:Bog_expansion} were set to zero in order to prevent the unphysical UV divergence in \eqref{eq:rho_TWA} from obscuring the physical results. With the bias, the system at the critical point remains stable against small fluctuations that add just a small quantum correction in \eqref{eq:rho_Bog}. The top panel demonstrates a perfect collapse interpolating between the predicted asymptotes.

The initial fluctuations in \eqref{eq:Bog_expansion} were included in the bottom panel of Fig. \ref{fig:collapse_M_c} showing magnetization averaged over random $\eta_n$. The UV-divergent fluctuations make plots depart from the collapsed plots in the top panel. 
The departure originates from the high-frequency Bogoliubov modes whose adiabaticity depends only on $\tau_Q$ while their mode eigenfunctions show a linear response to the bias.
Accordingly, for each bias, the departure begins at $\tau_Q$ that is independent of the bias and at a value of the order parameter that is proportional to $b$. 
Whereas at $\Omega_c$, the exact quantum fluctuations can be just neglected, in the following evolution below $\Omega_c$, long wavelength Bogoliubov modes trigger inhomogeneities that survive in the symmetry broken phase. The effect of the high-frequency fluctuations on the inhomogeneous pattern is averaged to zero on the time scale $\hat t$ that it takes the inhomogeneities to develop. In this respect, the high-frequency modes do not need to be truncated by hand.

The average order parameter is one of the characteristics that can probe the final state in the immiscible phase at the end of the ramp. Figure \ref{fig:collapse_M_f}(a) shows that final $M$ collapses in the non-adiabatic regime for small $b\tau_Q$. The collapse cannot extend to the complementary adiabatic regime, that is, for large $b\tau_Q$, because the order parameter saturates there at $1$ instead of remaining proportional to $\left(b/\Omega_c\right)^{1/3}$. At the end of the ramp, all particles end in the favourable component $1$. This does not preclude a collapse for a suitably modified scaling hypothesis.

One may notice that, in Fig. \ref{fig:collapse_M_f}(a), the asymptote $\propto (b\tau_Q)^{2/3}$ (valid for small $b\tau_Q$) crosses the saturation level $\propto b^{-1/3}$ achieved for large $b\tau_Q$ at $\tau_Q\propto b^{-3/2}$. Therefore, a simultaneous collapse in both regimes can be engineered by plotting unscaled order parameter $M$ as a function of $b^{3/2}\tau_Q$, see Fig. \ref{fig:collapse_M_f}(b). In the final state it is $\tau_Q\propto b^{-3/2}$, in place of $\tau_Q\propto b^{-1}$, that marks the actual crossover to the defect-free regime. In the next section, we will see the same crossover for the density of defects in the final state.

The final scaling is predicted by crossing the two asymptotes. The saturation level of the order parameter at $M=1$ for large enough $\tau_Q$ must be trivially true. The asymptote $\propto (b\tau_Q)^{2/3}$ for fast quenches is predicted by KZM within $\pm\hat t$ but it does not need to survive until the end of the ramp at $\Omega=0$. However, in first approximation, one can argue that a domain pattern that forms at $+\hat t$ survives until the end of the ramp and, therefore, the average order parameter determined by the proportion of the two immiscible phases survives as well. 

\begin{figure}[t!]
\vspace{-0cm}
\includegraphics[width=0.999\columnwidth,clip=true]{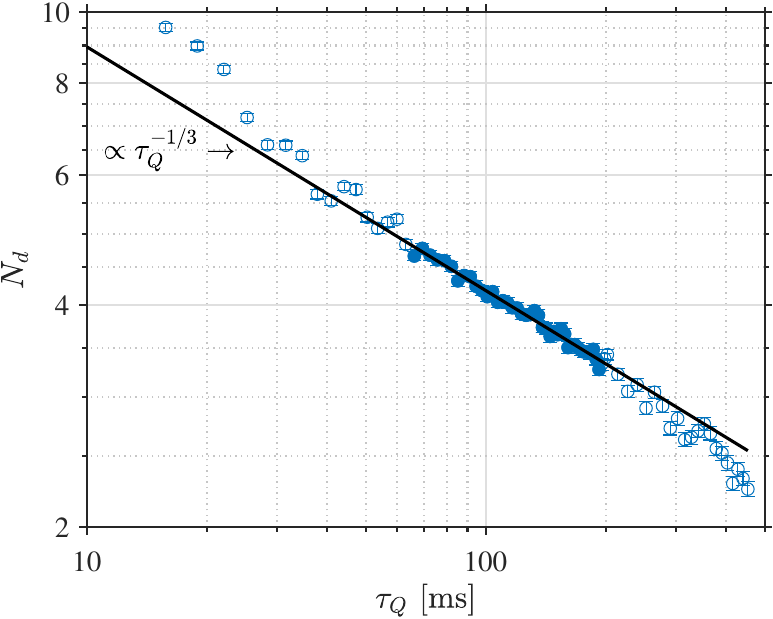}
\vspace{-0cm}
\caption{
{Defect density without bias.} 
Average number of defects as a function of $\tau_Q$. For intermediate quench times, the slope is $-1/3$ in consistency with $N_d\propto\hat\xi^{-1}$ with mean field exponents. With exact exponents, the slope $-1/2$ would be significantly different. 
For large $\tau_Q$, where the number of kinks goes down towards $2$, the curve begins to cross over to an exponential decay as the finite size of the system makes the transition adiabatic thanks to a finite gap in the spinon excitation spectrum. For small $\tau_Q$, kinks are overcounted as they are often difficult to distinguish from extra zero crossings due to strong fluctuations.
}
\label{fig:no_bias}
\end{figure}
\begin{figure}[t!]
\vspace{-0cm}
\includegraphics[width=0.999\columnwidth,clip=true]{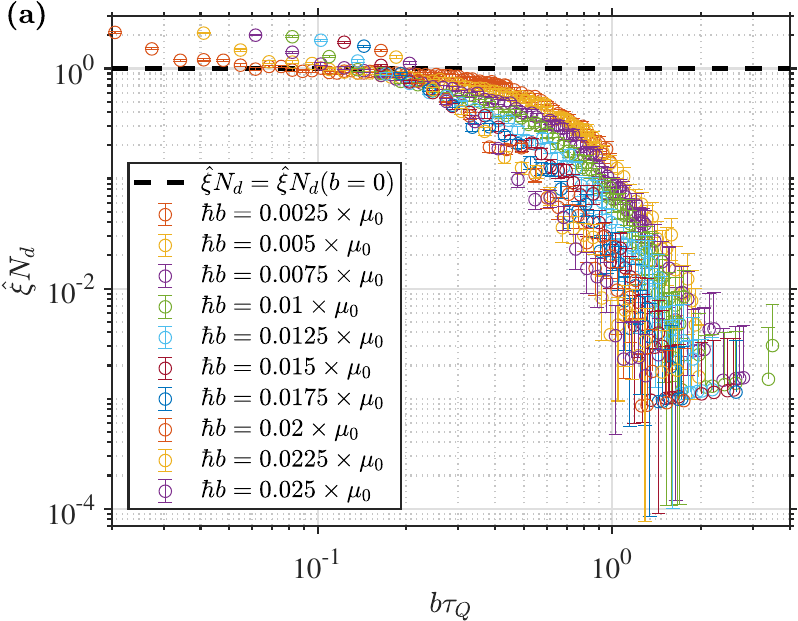}
\includegraphics[width=0.999\columnwidth,clip=true]{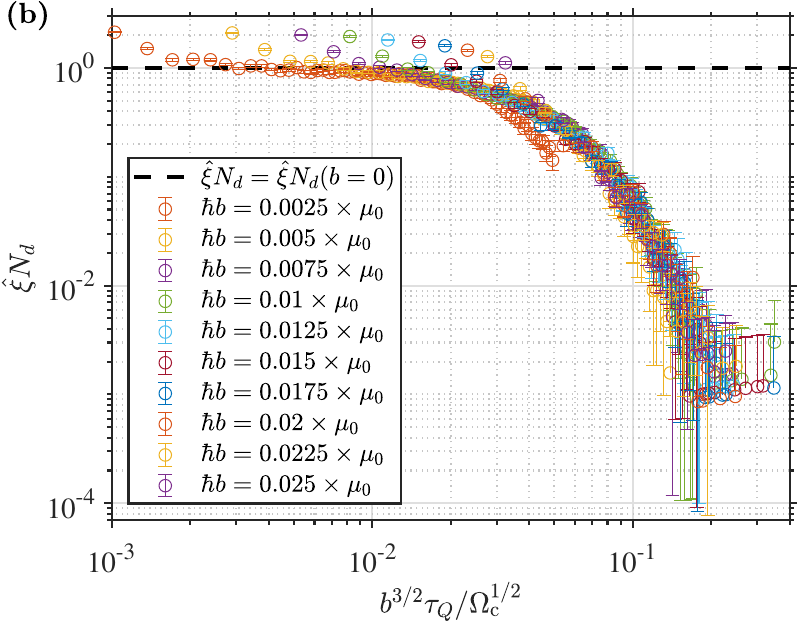}
\vspace{-0cm}
\caption{
{Defect density with bias.} 
In (a)
scaled number of defects as a function of $b \tau_Q^{\beta\delta/(1+z\nu)}$ for the mean-field critical exponents. The defects were counted deep in the immiscible phase. Their annihilation between $+\hat t$ and the counting, which is shown in Figs. \ref{fig:decimation1} and \ref{fig:decimation2}, explains why the collapse is not perfect.
In (b)
the same scaled defect density but as a function of $b^{3/2}\tau_Q$, similarly as in Fig. \ref{fig:collapse_M_f}, that is improving the collapse.
}
\label{fig:collapse_n}
\end{figure}

\section{ Density of kinks }
\label{sec:K}

The fluctuations in \eqref{eq:Bog_expansion} are essential for breaking the translational invariance and formation of kinks/defects separating domains of different immiscible phases. In the usual way\cite{QKZteor-r,KZexp-x}, one can argue that the density of defects, $n$, should satisfy a scaling hypothesis:
\be 
N_d = \hat\xi^{-1} F_N\left[ (t-t_c)/\hat\xi^z,  b \tau_Q^{\beta\delta/(1+z\nu)}  \right].
\label{eq:Fn}
\ee 
This scaling hypothesis is expected to hold in the KZ regime extending up to $\hat t$ where, unfortunately, counting defects is still obscured by relatively large fluctuations. If we want to avoid sophisticated filtering of the fluctuations, which would require extra theorizing and smuggling in some of the KZ assumptions, the counting has to be postponed until deep in the immiscible phase where the kinks have large magnitudes as compared to the quantum noise but where we can also anticipate some discrepancies with respect to the scaling hypothesis. 

We begin with zero bias, the case considered before in Refs. \onlinecite{sabbatiniPRL,sabbatiniNJP}, when defect density $N_d\propto \hat\xi^{-1}$ with a proportionality factor, $F_N\left[ (t-t_c)/\hat\xi^z,  0  \right]$, is dependent only on the scaled time. Numerical results deep in the immiscible phase are shown in Fig. \ref{fig:no_bias}. They demonstrate that $N_d\propto\hat\xi^{-1}\propto \tau_Q^{-1/3}$ is consistent with the data for the mean-field critical exponents and significantly different from $n\propto \tau_Q^{-1/2}$ predicted with the exact ones.  

\begin{figure}[t!]
\vspace{-0cm}
\includegraphics[width=0.999\columnwidth,clip=true]{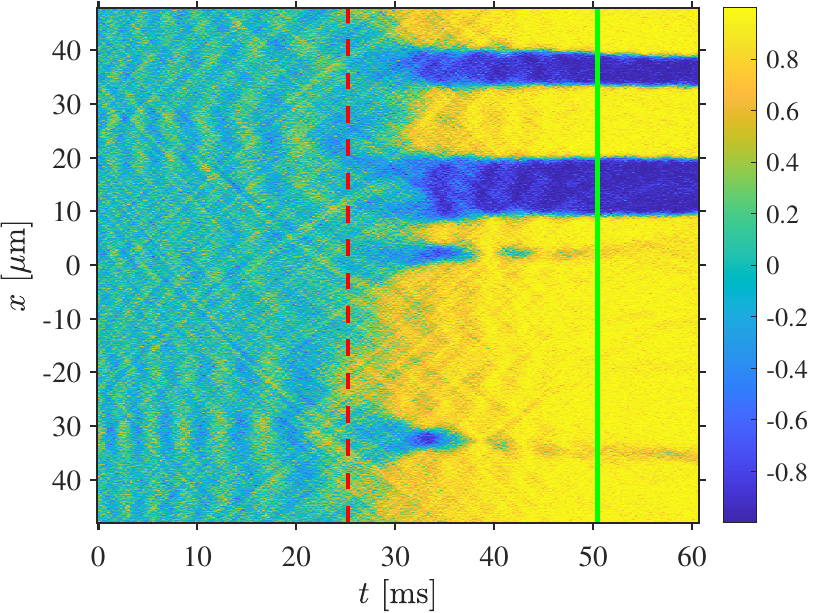}
\includegraphics[width=0.999\columnwidth,clip=true]{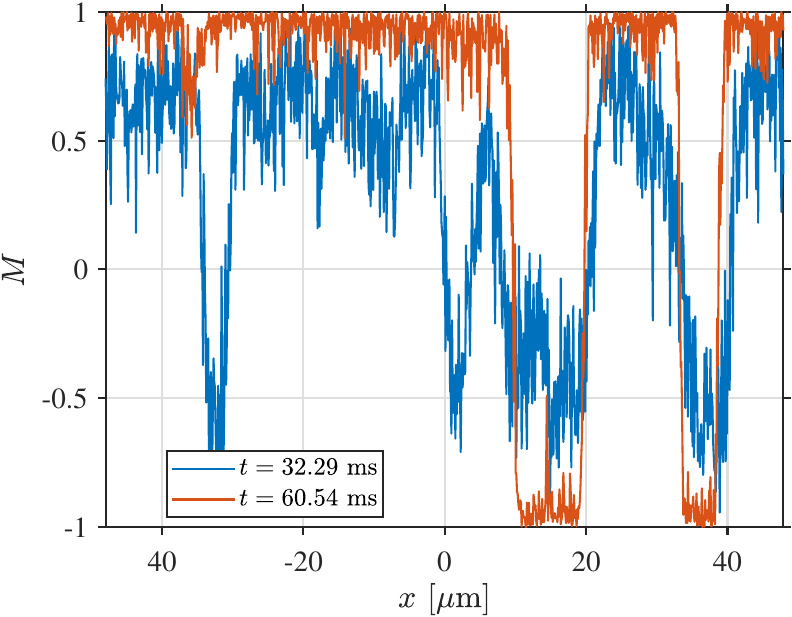}
\vspace{-0cm}
\caption{
{Defect annihilation. } 
The KZM predicts the density of defects at time $+\hat t$ immediately after the time evolution catches up with the ramp soon after crossing the critical point. These early defects can be too difficult to distinguish from quantum fluctuations to be reliably counted. Therefore, the actual counting is postponed until deep in the immiscible phase.
In the meantime, their number can be reduced by their mutual annihilation (or, equivalently, shrinking of the minority domains) as shown in the two panels where two domains disappear between $+\hat t$ and the end of the ramp.
}
\label{fig:decimation1}
\end{figure}

\begin{figure}[t!]
\vspace{-0cm}
\includegraphics[width=0.999\columnwidth,clip=true]{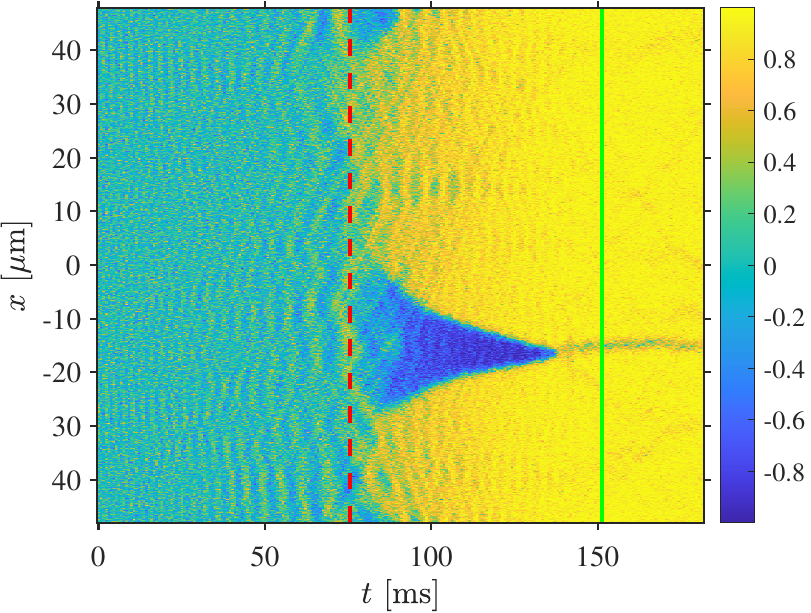}
\includegraphics[width=0.999\columnwidth,clip=true]{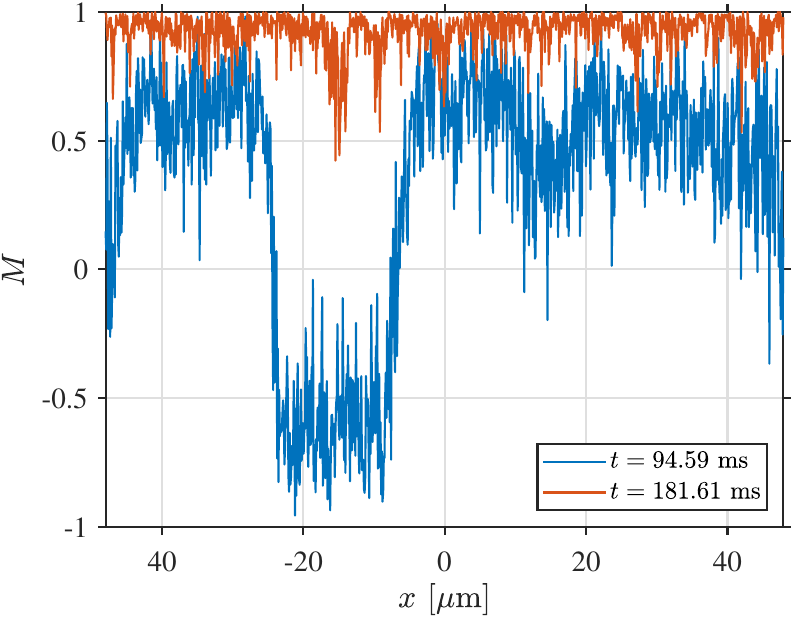}
\vspace{-0cm}
\caption{
{Defect annihilation. } 
Same as in Fig. \ref{fig:decimation1} but for a slower quench, deeper in the adiabatic regime. Here, a single domain shrinks and disappears between $+\hat t$ and the end of the ramp.
}
\label{fig:decimation2}
\end{figure}

For a weak bias, the scaling function in \eqref{eq:Fn} has two arguments. The second one, equal to $b\tau_Q$ for the mean-field exponents, discriminates between the nonadiabatic and the adiabatic regime for its small and large values, respectively. Without bias, the kinks are counted deep in the immiscible phase. Figure \ref{fig:collapse_n}(a) shows their scaled density as a function of $b \tau_Q$ for different bias strengths. Their collapse is not perfect, suggesting that with increasing bias, the final state becomes defect free for shorter $\tau_Q$ than suggested by the crossover value $b\tau_Q\approx1$.
The bias seems to suppress kinks not only by making the transition itself more adiabatic but also by favoring their annihilation between $+\hat t$ and the time of their counting deep in the immiscible phase. 

Indeed, examples of defect annihilation are shown in Figs. \ref{fig:decimation1} and \ref{fig:decimation2}. In both examples, a minority domain disappears together with its two delimiting kinks. 
In a similar way as for the final order parameter, and for the same reason, the collapse of the final kink density improves when scaled density $\hat\xi N_d$ is plotted as a function of $b^{3/2}\tau_Q$ instead of $b\tau_Q$, see Fig. \ref{fig:collapse_n}(b). 
The disappearance of small domains of the unfavourable phase reduces the number of kinks and, at the same time, brings the average order parameter closer to one.

\section{ Experimental feasibility  }
\label{sec:exp}

Two-component condensates have been experimentally realized using different atomic species \cite{mccarron2011,modugno2002}, atomic isotopes \cite{papp2008}, or spin states \cite{lin2011,tojo2010,nicklas2015b,cominotti2023}. The results presented in this paper correspond to a system of two strongly immiscible components with $\Delta \equiv g_{11}g_{22}/g_{12}^{2} = 0.25$. This corresponds to a spin healing length $\xi_\mathrm{s}$ that is relatively short, allowing us to identify the domains easily. In particular, the number of domains is obtained from our simulations by calculating the number of zero crossings of $M = (\rho_{1} - \rho_{2})/(\rho_{1} + \rho_{2})$. In experiments, $M$ can be easily extracted by performing  absorption imaging of the two components. The two hyperfine states are separated in energy by about 1000 times the linewidth of the optical transition used to probe them. Hence, one can take an absorption image of one component and immediately image the other with another light of a different frequency.

Specific to the results presented in this paper is the realization of phase separation with the same pair of atomic species using Feshbach resonances \cite{tojo2010}. No pair of hyperfine states of ${}^{87}$Rb and ${}^{23}$Na are naturally strongly immiscible, such as the case considered in this work. However, the combination of the $|F=1,m_\mathrm{F}=+1\rangle$ and $|F=2,m_\mathrm{F}=-1\rangle$ hyperfine states of ${}^{87}$Rb has an interspecies Feshbach resonance \cite{tojo2010,erhard2004} that can be tuned such that the two-component condensate is in the immiscible regime while keeping $g_{11} \approx g_{22}$. Nicklas \textit{et al.} \cite{nicklas2015b} have realized a binary quasi-one-dimensional condensate of ${}^{87}$Rb atoms with the above-mentioned hyperfine states and condition where $\Delta \approx 0.66$. It is also worth mentioning the possibility of using ${}^{23}$Na condensate, such as in a recent experiment of Cominotti \textit{et al.} \cite{cominotti2023} where they were able to realize a two-component condensate using the combination of the $|F=2,m_\mathrm{F}=-2\rangle$ and $|F=1,m_\mathrm{F}=-1\rangle$ hyperfine states of ${}^{23}$Na atoms, which in the absence of the coherent coupling is immiscible with $\Delta \approx 0.85$. It must be noted, however, that the use of a Feshbach resonance has a known disadvantage of inelastic atom losses \cite{inouye1998}, especially near resonance.

An alternative way to experimentally realize the miscible-immiscible phase transition of our system is via spin-orbit coupling of neutral atoms. In Ref. \onlinecite{lin2011}, the authors have coupled the two Zeeman sublevels of the $|F=1\rangle$ of ${}^{87}$Rb and were able to measure the phase separation of the dressed states across the critical point. The phase transition is achieved by ramping up the intensity of two slightly detuned lasers coupling the two hyperfine levels. This method has the advantage of reaching deeper into the immiscible regime without suffering atom losses, unlike the case for using a Feshbach resonance. However, as noted in Refs. \onlinecite{sabbatiniPRL,sabbatiniNJP}, the precise spatial arrangement of the dressed state could not be directly accessed. Instead, it was inferred from absorption imaging of the bare components. Consequently, while the increased separation and stability were advantageous, they necessitated a more complex detection process for determining the number of domains.

\section{ Conclusion }
\label{sec:conclusion}

This work unifies two themes in the theory of the quantum Kibble-Zurek mechanism (QKZM). One is the theory of the miscible-immiscible quantum phase transition in quasi-1D Bose-Einstein condensates developed in Refs. \onlinecite{sabbatiniPRL,sabbatiniNJP}. This mean-field quantum phase transition can be realized in binary condensate mixtures. The other is the QKZM with a bias that was investigated theoretically in Ref. \onlinecite{QKZteor-r} and whose classical version was experimentally verified in helium-3 \cite{KZexp-x}. The motivation for the study in Ref. \onlinecite{QKZteor-r} was to make the dynamics of the quantum phase transition adiabatic by applying a weak bias in order to speed up adiabatic quantum state preparation in a controlled way. Here, we propose to test this effect in a robust mean-field quantum transition.

We verified the QKZM scalings for the order parameter when the ramp is crossing the critical point but, as the system is further ramped into the immiscible phase, some defects are annihilated making the defect-free regime expand to faster non-adiabatic transitions. Phenomenological power laws were proposed to describe approximately the final order parameter and defect density. 

\vspace{6pt} 
The data used for the figures in this article are openly available from the GitHub repository at \onlinecite{GitHub_repo}.

\acknowledgements
Helpful discussions concerning experimental possibilities with Malcolm Boshier are gratefully acknowledged. 
This research was supported in part by the National Science Centre (NCN), Poland under project 2021/03/Y/ST2/00184 within the QuantERA II Programme that has received funding from the European Union Horizon 2020 research and innovation programme under Grant Agreement No 101017733 (F.B. and J.D.).
The research was also supported by a grant from the Priority Research Area DigiWorld under the Strategic Programme Excellence Initiative at Jagiellonian University (J.D.).
%

\bibliography{KZref.bib} 

\end{document}